\documentclass[12pt]{iopart}
\usepackage{iopams}  
\usepackage{setstack}
\usepackage{graphicx}
\usepackage[applemac]{inputenc}
\begin{document}
%------------------------------------------------------------------------------------------------------------------------------------------
\title{What is ``fundamental''?}
%------------------------------------------------------------------------------------------------------------------------------------------
\author{Matt Visser}
%------------------------------------------------------------------------------------------------------------------------------------------
\address{School of Mathematics and Statistics\\
Victoria University of Wellington, PO Box 600, \\
Wellington 6140, New Zealand}
%------------------------------------------------------------------------------------------------------------------------------------------
\ead{matt.visser@sms.vuw.ac.nz}
\begin{abstract}

\noindent
Our collective views regarding the question ``what is fundamental?'' are continually evolving. These ontological shifts in what we regard as fundamental are largely driven by theoretical advances (``what can we calculate?''), and experimental advances (``what can we measure?''). Rarely (in my view) is epistemology the fundamental driver; more commonly epistemology reacts (after a few decades) to what is going on in the theoretical and experimental zeitgeist.

foundations of science; ontology; epistemology; empiricalism.

\vskip 10 pt
\noindent 
{\sc Date}:   19 January 2018; 17 May 2018; \LaTeX-ed \today.

\vskip 10 pt
\noindent 
{\sc Purpose}:  Essay written for the FQXi 2018 essay contest: {\bf What is ``fundamental''?}

\end{abstract}

%\pacs{89.70.Cf;  89.70.-a}

\vspace{2pc}
\noindent

\maketitle

%%------------------------------------------------------------------------------------------------------------------------------------------
%%\bigskip
%%\hrule
%%\bigskip
%%------------------------------------------------------------------------------------------------------------------------------------------
%%\markboth{What is ``fundamental''?}{ }
%%\thispagestyle{empty}
%%\tableofcontents
%%\markboth{What is ``fundamental''?}{ }
%%------------------------------------------------------------------------------------------------------------------------------------------
%%\bigskip
%%\hrule
%%\bigskip
%%------------------------------------------------------------------------------------------------------------------------------------------
%%\clearpage
%%------------------------------------------------------------------------------------------------------------------------------------------
\markboth{What is ``fundamental''?}{}
%------------------------------------------------------------------------------------------------------------------------------------------
\def\d{{\mathrm{d}}}
\def\O{{\mathcal{O}}}
\def\omicron{o}
\def\CC{\mathbb{C}_0}
\def\N{\mathbb{N}}
\def\Z{\mathbb{Z}}
\def\Q{\mathbb{Q}}
\def\R{\mathbb{R}}
\def\C{\mathbb{C}}
%------------------------------------------------------------------------------------------------------------------------------------------
% leq 25000 characters
% body leq 9 pages
% references leq 1 page
%------------------------------------------------------------------------------------------------------------------------------------------
\parindent0pt
\parskip7pt

%\null
%\bigskip

Just what we view as ``fundamental'' is  a pragmatic decision based, at least partially, on what we can calculate and on what we can measure. (Sometimes, what we can \emph{easily} calculate and \emph{easily} measure.) Ultimately this is simply a reflection of the fact that science is a human endeavour, and sometimes certain topics are simply relegated to the ``too hard'' basket. This is not necessarily due to laziness, often is is a simple pragmatic decision based on what one might reasonably hope to  accomplish with finite resources.

\begin{quote}
For me, it is far better to grasp the universe as it really is than to persist in delusion, however satisfying and reassuring.\\
\null\hfill---Carl Sagan%, The Demon-Haunted World: Science as a Candle in the Dark
\end{quote}

In the 20th century ``fundamental physics'' was often viewed as being synonymous with special and general relativity, quantum physics, and relativistic quantum field theory.  This was a grossly simplistic, but exceedingly common, point of view. At least one quantum field theorist is infamous for referring to condensed matter physics as ``squalid state physics''; there were also quantum field theorists who viewed the entire geometrical interpretation of general relativity as irrelevant noise; and there was (and still is) considerable tension between some segments of the relativity and field theory communities.

\begin{quote} 
A university student attending lectures on general relativity in the morning and others on quantum mechanics in the afternoon might be forgiven for thinking that his professors are fools, or have neglected to communicate with each other for at least a century.\\
\null\hfill---Carlo Rovelli%, Seven Brief Lessons on Physics
\end{quote}

Foundations of quantum mechanix were for much of the 20th century often quietly ignored, when not being outright suppressed --- a common attitude being ``you will come to know the rules'', sometimes summarized as the ``shut up and calculate'' non-interpretation of quantum physics.  Indeed, as long as there were lots of viable ``bash and see'' scattering experiments in the pipeline, and relatively few viable experimental proposals for testing the ``collapse of the wavefunction'', then concentrating on what is achievable was not entirely irrational.  The subtlety then lies in recognizing when foundational experimental proposals are becoming viable; or when foundational theoretical approaches are becoming somewhat more mature and stable. 
Sometimes \emph{fiction} gives a good window into \emph{reality}: Consider the  \emph{fictional} musings of  the \emph{fictional} physicist Shevek~\cite{Le-Guin}:
\begin{quote}
...the physicists of [Einstein's] own world had turned away from
his effort and its failure, pursuing the magnificent incoherencies of
quantum theory, with its high technological yields, ... to arrive at a
dead end, a catastrophic failure of the imagination.\\
\null\hfill--- Shevek \emph{circa} 2500 CE
\end{quote}

The tension between quantum physics and classical general relativity is probably most acute when contemplating the black hole information ``puzzle''. (I point blank refuse to characterize this issue as a ``paradox''.) The tension arises only by extrapolating two extremely useful but approximate ``fundamental'' concepts (curved space quantum field theory and general relativistic classical event horizons) into realms where we really have no reason to trust them. 

\begin{quote}
God does not play dice with the universe.\\
\null\hfill---Albert Einstein%, The Born-Einstein Letters 1916-55
\end{quote} 

\begin{quote} 
Not only does God play dice but ... he sometimes throws them where they cannot be seen.\\
\null\hfill---Stephen Hawking
\end{quote}

What we view as ``fundamental astronomy'' has also shifted radically over the course of the 20th century. The idea that nuclear reactions could power stellar luminosity was new and radical in the 1929s; cosmology and cosmography (as science) did not really exist until the 1920s. The ``big bang'' was not widely accepted until the 1950s.  

\begin{quote}
Don't let me catch anyone talking about the universe in my department.\\
\null\hfill---Ernest Rutherford
\end{quote}

Even special relativistic astrophysics was largely a post-1960s phenomenon.  This aspect of the evolution of the cosmological zeitgeist was primarily driven by an interplay between improved observational and improved calculational techniques.

\begin{quote}
Far out in the uncharted backwaters of the unfashionable end of the western spiral arm of the galaxy lies a small unregarded yellow sun. Orbiting this at a distance of roughly ninety-two million miles is an utterly insignificant little blue green planet whose ape-descended life forms are so amazingly primitive that they still think digital watches are a pretty neat idea. \\
\null\hfill---Douglas Adams%, The Hitchhiker's Guide to the Galaxy
\end{quote}

The more things change, the more they stay the same --- the now 20 year old acceptance of the accelerating universe implies, (though no-one would be so crass as to actually say so), that the 1950s notion of the ``heat death of the universe'' is back with a vengeance ---  though to make it look more respectable and up-to-date, maybe we should now re-label  it as the ``entropy death'':

\begin{quote}
The bright sun was extinguish'd, and the stars\\
Did wander darkling in the eternal space.\\
\null\hfill---Byron%George Gordon Byron
\end{quote}

At one stage the ``geocentric'' point of view was considered fundamental, no longer:

\begin{quote}
We have developed from the geocentric cosmologies of Ptolemy and his forebears, through the heliocentric cosmology of Copernicus and Galileo, to the modern picture in which the earth is a medium-sized planet orbiting around an average star in the outer suburbs of an ordinary spiral galaxy, which is itself only one of about a million million galaxies in the observable universe.\\
\null\hfill---Stephen Hawking%, A Brief History of Time
\end{quote}

In mathematics perhaps the biggest shift in mathematical fundamentals in the 20th century was initiated by the G\"odel incompleteness results --- effectively demonstrating the limitations of the axiomatic framework: 

\begin{quote}
Kurt G\"odel ... proved that the world of pure mathematics is inexhaustible. No finite set of axioms and rules of inference can ever encompass the whole of mathematics. Given any finite set of axioms, we can find meaningful mathematical questions which the axioms leave unanswered. This discovery ... came at first as an unwelcome shock to many mathematicians. It destroyed ... the hope that they could solve the problem of deciding by a systematic procedure the truth or falsehood of any mathematical statement. ... G\"odel's theorem, in denying ... the possibility of a universal algorithm to settle all questions, gave ... instead, a guarantee that mathematics can never die. ... there will always be, thanks to G\"odel, fresh questions to ask and fresh ideas to discover.\\
\null\hfill---Freeman Dyson%, Infinite in All Directions (1988)
\end{quote}

This in turn connects back to Wigner's  ``unreasonable effectiveness of mathematics'' when applied to physics (or any other scientific endeavour for that matter):

\begin{quote}
Philosophy [nature] is written in that great book which ever is before our eyes --- I mean the universe --- but we cannot understand it if we do not first learn the language and grasp the symbols in which it is written. The book is written in mathematical language, and the symbols are triangles, circles and other geometrical figures, without whose help it is impossible to comprehend a single word of it; without which one wanders in vain through a dark labyrinth.
\null\hfill---Galileo Galilei
\end{quote} 

\enlargethispage{20pt}

\clearpage
\begin{quote}
There cannot be a language more universal and more simple, more free from errors and obscurities ... more worthy to express the invariable relations of all natural things [than mathematics]. [It interprets] all phenomena by the same language, as if to attest the unity and simplicity of the plan of the universe, and to make still more evident that unchangeable order which presides over all natural causes.\\
\null\hfill---Joseph Fourier%, The Analytical Theory of Heat
\end{quote} 

\begin{quote}
As far as the laws of mathematics refer to reality, they are not certain;\\ 
and as far as they are certain, they do not refer to reality.\\ \null\hfill---Albert Einstein
\end{quote}

In biology, the mid-century revolution in molecular biology (specifically the discovery of RNA and DNA) completely rewrote our ideas of what was fundamental biology --- with ``reverse engineering'' of the genetic code eventually leading to the development of bio-informatics as an almost-stand-alone discipline bridging biology, mathematics, and computer science. That there are still fundamental and foundational issues at play here can be seen from the relative ``vagueness'' of the definition of a gene --- if a segment of genetic code seems to do something useful, we will call it a gene.  This definition somehow lacks the ``precision'' that most mathematicians or theoretical physicists would be comfortable with, but at least for the time being, it is certainly pragmatically useful.

\begin{quote}
The beauty in the genome is of course that it's so small. The human genome is only on the order of a gigabyte of data ... which is a tiny little database. If you take the entire living biosphere, that's the assemblage of 20 million species or so that constitute all the living creatures on the planet, and you have a genome for every species the total is still about one petabyte, that's a million gigabytes --- that's still very small compared with Google or the Wikipedia and it's a database that you can easily put in a small room, easily transmit from one place to another. And somehow mother nature manages to create this incredible biosphere, to create this incredibly rich environment of animals and plants with this amazingly small amount of data.\\
\null\hfill---Freeman Dyson
\end{quote}

One particularly difficult issue when working on foundations or fundamentals in any field is that one quickly encounters the crackpot fringe; the noise/signal ratio is often distressingly high. A key point to keep firmly in mind is that (crackpot) $\neq$ (wrong). There are crackpots who are (often accidentally) correct, and there are people who are wrong on some issue without being in any way crackpot. Detecting a crackpot has more to do with the style (and internal logical incoherence) of the arguments being used, than the actual technical points being raised. 

\begin{quote}
How is it that hardly any major religion has looked at science and concluded, ``This is better than we thought! The universe is much bigger than our prophets said, grander, more subtle, more elegant?'' Instead they say, ``No, no, no! My god is a little god, and I want him to stay that way.'' 
%A religion, old or new, that stressed the magnificence of the Universe as revealed by modern science might be able to draw forth reserves of reverence and awe hardly tapped by the conventional faiths.
\\
\null\hfill---Carl Sagan%, Pale Blue Dot: A Vision of the Human Future in Space
\end{quote}

While our views on what is fundamental change with the times --- this should not be taken as an excuse for not doing the hard work. It is distressingly easy to redefine any problem away using linguistic tricks --- one should be very careful to make sure one is doing more than just playing linguistic tricks. Developing a formalism that is so badly designed and implemented that one cannot even formulate the question you were originally interested in does not mean you have actually solved the question. There is a long-standing joke in the theoretical physics community regarding the ``redefinition group'' --- like the ``renormalization group'' it is actually a semi-group --- the flow is one-way and inverse operations need not necessarily exist. In applying the ``redefinition group'' one simply successively redefines concepts and notions until the answer you have turns into the answer you want. 

For instance, some segments of the theoretical physics community are now greatly enamoured of ``post-empirical science'', a quite remarkable oxymoron. In fact, if one looks more carefully at what the philosophers of science are actually saying, they seem to be arguing for ``semi-empirical theory verification'' --- if you cannot directly test the theory of interest, then look around the edges, find some ``parallel'' theories (models) slightly different from what you are really interested in, but close enough that there is at least some overlap, and then empirically test those ``parallel'' models. Semi-empirical techniques along these lines then still tie into Popper's notion of ``falsifiability''. I would argue that ``falsifiability'' should be an aspiration, not necessarily the be-all and end-all of the definition of science. There is nothing wrong with a temporary tactical retreat from falsifiability, while one is putting one's house in order --- but the retreat should be both temporary and tactical, not an open-ended admission of defeat.  
(If the temporary tactical retreat takes up a significant fraction of one's professional lifetime, then properly speaking it is neither tactical not temporary,  one should re-think the matter.)

 In summary, simply because science is a human endeavour, the precise details of what we consider to be fundamental will unavoidably change with the times. On this topic at least, it is perhaps safer to stick to general themes, such as the quantum-to-gravity interface, and avoid excessive precision.

\bigskip
%\begin{quote}
%For me, it is far better to grasp the Universe as it really is than to persist in delusion, however satisfying and reassuring.\\
%\null\hfill---Carl Sagan%, The Demon-Haunted World: Science as a Candle in the Dark
%\end{quote} 

%\begin{quote} 
%How can one look happy when he is contemplating the anomalous Zeeman effect?\\
%\null\hfill--- Wolfgang Pauli
%\end{quote} 

%\begin{quote} 
%Another very good test some readers may want to look up, which we do not have space to describe here, is the Casimir effect, where forces between metal plates in empty space are modified by the presence of virtual particles.
%\end{quote} 
%
%\begin{quote} 
%Thus virtual particles are indeed real and have observable effects that physicists have devised ways of measuring. Their properties and consequences are well established and well understood consequences of quantum mechanics.? 
%? Gordon L. Kane
%\end{quote} 

\medskip
\centerline{---\,\#\#\#\,---}

\clearpage
%--------------------------------------------------------------------------------------------------------------------------
\ack
%--------------------------------------------------------------------------------------------------------------------------

Supported via the Marsden Fund,  administered by the Royal Society of New Zealand.

%\clearpage
%------------------------------------------------------------------------------------------------------------------------------------------
\section*{Some reading}
%------------------------------------------------------------------------------------------------------------------------------------------

%------------------------------------------------------------------------------------------------------------------------------------------

\begin{thebibliography}{69}
\bibitem{anyone}
  ``Has anyone seen that bastard Schr\"odinger?"
\end{thebibliography}
\end{document}